\documentclass[12pt]{article}
\usepackage[dvips]{graphicx}

\begin{document}

\title{Stochastic effects in
the growth of droplets}

\author{V.Kurasov}

\date{viccur@bk.ru }

\maketitle

\begin{abstract}
The effects of stochastic absorption and ejection
of molecules by growing droplets have been
considered. Both analytical and numerical
approaches have been used. They demonstrate the
satisfactory coincidence. It is proved that in
general case corresponding to the asymptotic at
big numbers of molecules  in the critical embryo
the effects of stochastic growth are small in
comparison with the effects of stochastic
appearance of droplets.
\end{abstract}

This paper  continues the set of publications \cite{lanl1},
\cite{lanl2}, \cite{lanl3}, \cite{lanl4} about  stochastic effects
in nucleation.  It is known that the growth of embryos occurs in
nucleation kinetics stochastically. Really, in a given moment of
time $t$ the probability to absorb a molecule to a given cluster of
a size $\nu$ (i.e. containing $\nu$ molecules of a liquid phase)
from the time $t$ up to the time $t+dt$
 is $W^+ dt$ where $dt$ is an
elementary interval and $W^+$ is the absorption
coefficient calculated in the free molecular
regime of substance exchange as
$$
W^+ = \frac{1}{4} v_t \rho S \alpha
$$
Here $v_t$ is the thermal velocity of motion of
the molecule of vapor, $\rho$ is the density of
vapor, $S$ is the surface of the cluster,
$\alpha$ is the condensation coefficient. This
formula can be easily derived from the simple gas
kinetic theory.

The evident restriction for $dt$ is $dt \ll
(W^+)^{-1}$.

There exists an analogous probability to eject a
molecule into a vapor phase $W^- dt$ where $W^-$
is the ejection coefficient. Ordinary one can put
$W^-$ equal to $W^+$ at the density of vapor
equal to the saturated  vapor over the curved
surface of the embryo. Precisely speaking this is
one of the postulates of nucleation theory
stating that the internal state of embryo does
not strongly depend on the state of vapor. So,
$$
W^- = W^+ |_{\rho=\rho_e(\nu)}
$$
where $\rho_e$ is the density of vapor in equilibrium with  the
embryo of the size $\nu$.

Here the additional restriction for $dt$ is $dt
\ll (W^-)^{-1}$.

Since the most frequent ejections and absorptions
take place for big embryos one can check
restrictions for the "maximal" droplet in the
system. Certainly,  this is the supercritical
droplet and here $W^- < W^+$. So, one can check
only $dt \ll (W^+)^{-1}$.

So, we can see that the process of growth is
principally stochastic and required the
corresponding description.

From the first point of view to use the law of regular growth the
following arguments are presented:
\begin{itemize}

\item
The characteristic size of droplets is rather
big. So, the characteristic dispersion which is
about the square root of the total sum of both
absorption and ejection events is enough small in
comparison with the mean number of molecules
inside the cluster.

\item
Ordinary the number of droplets is so giant that
the averaging over the whole volume of the system
leads to the compensation of deviations of
particular droplets from the mean value of the
size coordinate.

\end{itemize}

Both these remarks explain the negligible
character of stochastic corrections in droplets
growth and we have to discuss them.

About the first remark one can come to the
following notes:
\begin{itemize}
\item
It is well known that the rate of embryos growth has the avalanche
character (see \cite{TMF}). Namely the avalanche consumption leads
to the fact that the dispersion $\delta \nu$ of the number of
molecules $\nu$ inside the droplet doesn't satisfy the ordinary
relation $\delta \nu \sim \nu^{1/2}$. Here, in the free molecular
regime of growth we shall show that $\delta \nu $ approximately has
another power  behavior $\delta \nu \sim \nu^{1- \epsilon}$ with
$\epsilon$ close to $0$.

So, the relative weight of fluctuations doesn't
disappear rapidly. Fortunately, $\epsilon $ is
positive and the formal convergence takes place.

\item
It is quite possible that due to the complex dependence of $W^+,
W^-$ on $\nu$  the mean value $<\nu>$ of the droplet size $\nu$
averaged over many  attempts differs from the value $\nu_{reg}$
calculated on the base of regular law of droplets growth.

\end{itemize}

So, the special consideration is necessary. Fortunately, in some
sense (it will be clear later) one can show that $\nu_{reg}$ is
close to $<\nu>$.

The second remark  faces the following notes:
\begin{itemize}
\item
 The observed number of droplets in the system
isn't giant, this specific feature  has been fully explained in
investigation of effects of stochastic appearance of supercritical
clusters \cite{lanl1}.
\item
 Even being averaged over the giant
number of droplets the mean coordinate can differ from the value
predicted on the base of the regular rate of growth.  The reason
 can be the  non-linear dependence of $\nu(t)$ on $t$.
\end{itemize}

 Here the role of only discrete values of $\nu$
 is
also important, but even with account of discrete
character of $\nu$ one can not come to
coincidence of approach based on the regular
growth and approach of stochastic growth. Here
the difference is very small but still it exists.

\section{The model}

The complex dependence of $W^-$ on $\nu$ occurs
mainly through the dependence of $\rho_{e}$ on
$\nu$. But one can not get precise final results
in analytical form  taking into account namely
this dependence without extracting asymptotes.

For the supercritical clusters one can use instead of the mentioned
density the density of the  vapor saturated over the plane surface.
Then
$$
W^- \approx W^-_0  = W^+ |_{\rho=\rho_e(\infty)}
$$
where $\rho_e(\infty)$ is the density of the
saturated vapor over the plane surface.

In further analysis we shall use $W^-_0$ instead
of $W^-$.

In renormalized scale of time which we choose for
simplicity one can write that
$$
W^-_0  = 3 \nu^{2/3}
$$
$$
W^+ = 3 (\zeta+1) \nu^{2/3}
$$
Here $\zeta$ is the supersaturation of the vapor
$$
\zeta = \rho/\rho(\infty) -1
$$

The regular law growth for the supercritical cluster in this time
scale can be written as
$$
\frac{d\nu}{dt} = W^+ - W^-_0  = 3 \zeta
\nu^{2/3}
$$
Here this law is written already for the
supercritical droplets.

Stochastic model used in numerical simulation
will be the following:
\begin{itemize}
\item
At initial moment of time $t=0$ the droplet is
situated at $\nu= \nu_0$
\item
At every step $dt$ two random numbers $r_+$ and
$r_-$ are generated. If $r_+ < W_+ dt$ then $\nu$
goes to $\nu+1$, if $r_2 < W_- dt$ then $\nu$
goes to $\nu -1$.
\item
At $t=t_{fin}$ the process of growth stops and
the attained value $\nu_{fin}$ will be the result
of calculations
\end{itemize}

Several (many) attempts  have been made and $<\nu>$ is calculated as
$$
<\nu> = \frac{\sum \nu_{fin}}{N}
$$
where $N$ is the number of attempts. The value of characteristic
fluctuation $\delta \nu$ will be calculated as\footnote{The value
$<\nu>$ here is calculated on the base  of the separate set of
attempts.}
$$
\delta \nu = \sqrt{2 ( \frac{\sum \nu_{fin}^2}{N}
- <\nu>^2)}
$$

One has to stress that it is impossible to put
$\nu_0$ close to $0$ because we consider the
supercritical droplets. Moreover  when $\nu_0$
goes to $0$ one can expect divergence of $\delta
\nu$. Fortunately this divergence is rather weak.
It will be discussed later.

Another reason to forbid the small values of
$\nu_0$ is that in this process one can not
attain $\nu=0$ in principle. What shall we do
with the totally dissolved embryo? This is the
necessary disadvantage of the approach used here.
So, one has to take $\nu_0$ greater than $10$.
The last value ensured the necessary accuracy and
negligible intensity  of the process of
dissolution.

\section{Example for the mean coordinate}

Let the characteristic size of the cluster be
$\nu_{fin} = 1000$ molecules. Let the
supersaturation be $\zeta = 4$ (during the
nucleation period it falls but not so
essentially, so one can take the characteristic
value).

We start at $\nu_{0}=20$ and try to attain
$\nu_{fin} = 1000$. We calculate the time
necessary to attain it according to macroscopic
regular law in continuous approximation
$$
t_{cont} = \frac{\nu_{fin}^{1/3} -
\nu_{st}^{1/3}}{\zeta}
$$
(in corresponding time units) and get $t_{cont} = 1.8214 $. Then we
recalculate the time according to regular law with discrete jumps $
\zeta t_{disc} = \frac{1}{3} \sum_{i=\nu_{st}}^{i=\nu_{fin}-1}
i^{-2/3} $ and get $t_{disc} = 1.8267    $.

The  value $t_{cont}$  has to be corrected up to the integer number
of elementary time intervals $dt$. In this example $dt = 0.00002$.
Then $\nu_{fin}$ will be slightly another.

It is more profitable to fulfil the simple
simulation of regular growth. At every step $dt$
instead of $\nu$ we get $\nu+ 3 \zeta \nu^{2/3}
dt$. The final value will be marked as
$\nu_{ff}$. This way allows to estimate the
deviation caused by the finite value of $dt$.

The general result of simulations is that  the value $\nu_{ff}$ is
practically the same as $<\nu>$.

To note the difference between these values  one has to see that
\begin{equation}\label{tty}
|<\nu> - \nu_{ff}| \gg \delta \nu / \sqrt{N}
\end{equation}
So, we have to calculate $\delta \nu$.

Now we turn to the dispersion. The mean number of events  can be
estimated as
$$
<E_{tot}> = \frac{(\zeta+2)}{\zeta} \nu_{fin}
$$
So, the estimate for the mean standard deviation of the size $\nu$
from the mean value $<\nu>$ in a particular attempt  will be
$$
\delta \nu = \sqrt{2 <E_{tot}>}
$$
Namely,
$$
\delta \nu \approx \sqrt{2000} \approx  45
$$

Numerical simulations give $ \delta_{cal} \nu \approx 155 $.

The discrepancy  will be a matter of discussions.

For the final values we have the following
results
$$
\nu_{ff} = 1004
$$
$$
<\nu> = 998
$$
To ensure that
$$
 <\nu> \neq \nu_{ff} \neq \nu_{fin} = 1000
$$
we fulfilled $10000$ attempts. Formally the necessary inequalities
(\ref{tty}) are satisfied. But still this  discrepancy can be the
result of small errors in simulation. Really, the random number
generators are not so perfect. Also there has to be a special
correction caused by the prescribed sequence of possible adsorption
and ejection events. This weak correlation remained without our
attention.

Certainly, the deviation in the mean coordinate
is out of practical interest. It is much more
interesting to consider the deviation in
dispersion.

\section{Analytical explanation for the mean value}

To see that $<\nu> \approx \nu_{ff}$ we start
from the master equation in Fokker-Planck
approximation for the partial distribution
function $n$ over size $\nu$:
$$
\frac{\partial n}{ \partial t} =
\frac{\partial}{\partial \nu} \large[ W^+ n^e
\frac{\partial}{\partial \nu} \frac{n}{n^e}
\large]
$$
where $n^e$ is the formal equilibrium distribution. The last
equation is valid for small values of supersaturation.

From the equation of detailed balance it follows
that
$$
W^-(\nu+1) n^e(\nu+1) = W^+ (\nu) n^e(\nu)
$$
Since
$$
n^e \sim \exp(-F(\nu))
$$
where $F$ is the free energy of the embryo, one
can get explicit kinetic equation on $n$.

In the cappilary approximation
$$
F = -\ln(\zeta+1) \nu + a \nu^{2/3}
$$
where $a$ is a scaled surface tension.

 For the supercritical
embryos
$$
\partial F / \partial \nu
\approx - \ln(\zeta+1)
$$

Taking into account that $ W^+ = 3 (\zeta+1)
\nu^{2/3} /\tau $, $ W^- = 3  \nu^{2/3} /\tau $
and acting in a scale where $3/\tau = 1$ one can
get
\begin{equation}\label{pp}
\frac{\partial n}{\partial t} = - \frac{\partial }{\partial \nu}(
\zeta n  \nu^{2/3} - \frac{\zeta + 2}{2}  \frac{\partial }{\partial
\nu} n  \nu^{2/3} )
\end{equation}

The next step is to note that asymptotically at $\nu \rightarrow
\infty$
\begin{equation}\label{rr}
\frac{\partial }{\partial \nu}  (n \zeta  \nu^{2/3}  - \frac{\zeta +
2}{2} \frac{\partial }{\partial \nu}  \nu^{2/3} n) \approx \nu^{2/3}
\frac{\partial }{\partial \nu} (n \zeta - \frac{\zeta + 2}{2}
\frac{\partial }{\partial \nu} n)
\end{equation}

Since the initial condition is $n \sim \delta(\nu - \nu_0)$ one can
suppose that
$$
\frac{\partial}{\partial \nu}  \nu^{2/3}  n \approx
 \nu^{2/3} \frac{\partial}{\partial \nu}    n
$$
Then the kinetic equation (\ref{pp}) can be written as
$$
\frac{\partial n}{\partial t} = -\nu^{2/3}
\frac{\partial }{\partial \nu} ( \zeta n -
 \frac{\zeta
+ 2}{2} \frac{\partial }{\partial \nu} n ) -
\frac{2}{3} \nu^{-1/3} ( \zeta n -  \frac{\zeta +
2}{2} \frac{\partial }{\partial \nu} n )
$$

The second term
$$
II =- \frac{2}{3} \nu^{-1/3} ( \zeta n -
 \frac{\zeta
+ 2}{2} \frac{\partial }{\partial \nu} n )
$$
can be comparable with the fist term
$$
I = -\nu^{2/3} \frac{\partial }{\partial \nu} (
\zeta n -  \frac{\zeta + 2}{2} \frac{\partial
}{\partial \nu} n )
$$
only when
$$
| \frac{\partial }{\partial \nu} ( \zeta n - \frac{\zeta + 2}{2}
\frac{\partial }{\partial \nu} n )| \simeq
 \nu^{-1}
| \zeta n -  \frac{\zeta + 2}{2} \frac{\partial }{\partial \nu} n |
$$
which means that we are near the stationary
solution corresponding to
$$
( \zeta n -  \frac{\zeta + 2}{2} \frac{\partial
}{\partial \nu} n ) = J_s = const
$$
where $J_s$ is the stationary flow. But here we consider the growth
of a single droplet, the situation is opposite and the true initial
condition is
$$
n(\nu_0) \sim \delta (\nu- \nu_0)
$$
So, here the second term can be neglected and the
relation (\ref{rr}) can be justified.

Now we can see that having neglected $II$ in
kinetic equation one can reduce it to
\begin{equation}\label{r}
\frac{\partial n}{\partial t} = - \frac{1}{3}
\frac{\partial }{\partial r} ( \zeta n -
\frac{\zeta + 2}{2} \frac{1}{3r^2} \frac{\partial
}{\partial r} n )
\end{equation}
for $r = \nu^{1/3}$.

We see that the last equation is the ordinary
diffusion equation with constant velocity  of
regular growth and the coefficient $D_{r}$ of
diffusion along $r$-axis written as
\begin{equation}\label{difco}
D_{r} =\frac{1}{9r^2} \frac{\zeta + 2}{2}
\end{equation}
Since $D_{r}$ is a decreasing function, one can see that the
distribution function over $\nu$ (as a function of $r$) will be a
well localized function in $r$-scale (and, hence, in $\nu$-scale).
So, the characteristic relative width of $n$ will be small.

This conclusion will also lead to the self
consistency of negligible character of the second
term in kinetic equation.

Under the constant value of $D_{r}$ the
characteristic width $\delta r$ will have an
order $r^{1/2}$ and the characteristic value of
$r$ is many times greater than the characteristic
width. Here the situation leads to a more strong
inequality.

To see the effects of stochastic growth one can
also calculate the derivative $<\frac{d}{dt} \nu
>$. The expression is the following
$$
<\frac{d}{dt} \nu > = \frac{d}{dt}
\int_0^{\infty}  \nu n(\nu,t) d\nu
$$
Then
$$
<\frac{d}{dt} \nu > = - \int_0^{\infty} d\nu  \nu
 \frac{\partial}{\partial \nu} \nu^{2/3}( n \zeta -
\frac{\zeta + 2}{2} \frac{\partial n}{\partial \nu} )
$$
or
$$
<\frac{d}{dt} \nu > =  - \int_0^{\infty} n \zeta \nu^{2/3} d \nu +
\frac{\zeta + 2}{2} \int_0^{\infty} d\nu \nu^{2/3}\frac{\partial
n}{\partial \nu}
$$
Note that we have to use namely precise equation
(\ref{pp}) without transformation (\ref{rr}). If
we use (\ref{rr}) we get the additional wrong
factor $5/3$.

The first integral
$$
I_1 = \int_0^{\infty} n \zeta \nu^{2/3} d \nu
$$
corresponds to the regular growth and the second
integral
$$
I_2 = \frac{\zeta + 2}{2} \int_0^{\infty} d\nu
\nu^{2/3}\frac{\partial n}{\partial \nu}
$$
represents corrections.

 One can see that integration by parts gives
$$
I_2 =  \frac{2}{3}  \frac{\zeta + 2}{2} \int_0^{\infty}  n
\nu^{-1/3} d\nu
$$
Since $n$ is well localized in a relatively small
region $I_2$ can be presented as
$$
I_2 \sim   \frac{2}{3}  \frac{\zeta + 2}{2} <\nu^{-1/3}>
$$
and it is rather small in comparison with $I_1$
which can be estimated as
$$
I_1  \sim \zeta <\nu^{2/3}>
$$

So, the smallness of correction to the mean coordinate is proven
analytically.

One has to note that the mentioned conclusion is
rather typical for the  processes of such kind.
We can make our arguments more strong if we
recall the Uhlenbeck process. One can note that
for the Uhlenbeck process
$$
\frac{
\partial n }{\partial t}
= \gamma \frac{\partial}{\partial x } (xn) + D
\frac{\partial^2}{\partial x^2} n
$$
with two parameters $\gamma$ and $D$
the Green's function is known
and has the exponent form:
$$
n(x,t|x',t') = \sqrt{\frac{\gamma}{2 \pi D
(1-\exp(-2\gamma (t-t')))}} \exp(-\frac{\gamma
(x-\exp(-2\gamma (t-t')x'))^2} {
2D(1-\exp(-2\gamma (t-t'))) })
$$
This exponent form lies in correspondence with
the absence of correction to the regular growth.
Really,
$$
\int \frac{dx}{dt} n(x) dx = \gamma \int x n(x)
dx
$$
since $dx/dt \sim \gamma  x$. Then
$$
\int x n(x) dx = x_0 + \int(x-x_0) n(x) dx
$$
with arbitrary $x_0$.

The coordinate of the regular growth  $x_0$ will
be $\exp(-2\gamma (t-t')x')$. Then the Green's
function is the Gaussian symmetrical on $x-x_0$.
Then $<x> = x_0$. As the result the first term
$x_0 = dx_0 /dt$ describes the regular evolution
and the second term $\int(x-x_0) n(x) dx$ is
absent. This is one of the   reasons why the
Green's function is known here.

In our process the analogous property is absent. But the last
example shows that at least the effects of the difference between
$d<x>/dt$ and $<dx/dt>$ are moderate because in situations $dx/dt
\sim const$ and $dx/dt \sim x$ (our situation lies between these
laws) they are absent.

\section{Numerical results for dispersion
as a function of the final size}

The next task is to calculate dispersion $\delta
\nu$ as a function of $\nu_{fin}$ and $\nu_0$.
Here there is no need to average over such a
giant number of attempts, it is quite sufficient
to take $200$ attempts to get suitable result for
dispersion. The results of simulation are the
following:
\begin{itemize}
\item
For $\nu_0 = 20$ one can see in the Figure 1 the
following dependence of $\delta \nu$ on
$\nu_{fin}$

\begin{figure}[hgh]

\includegraphics[angle=270,totalheight=8cm]{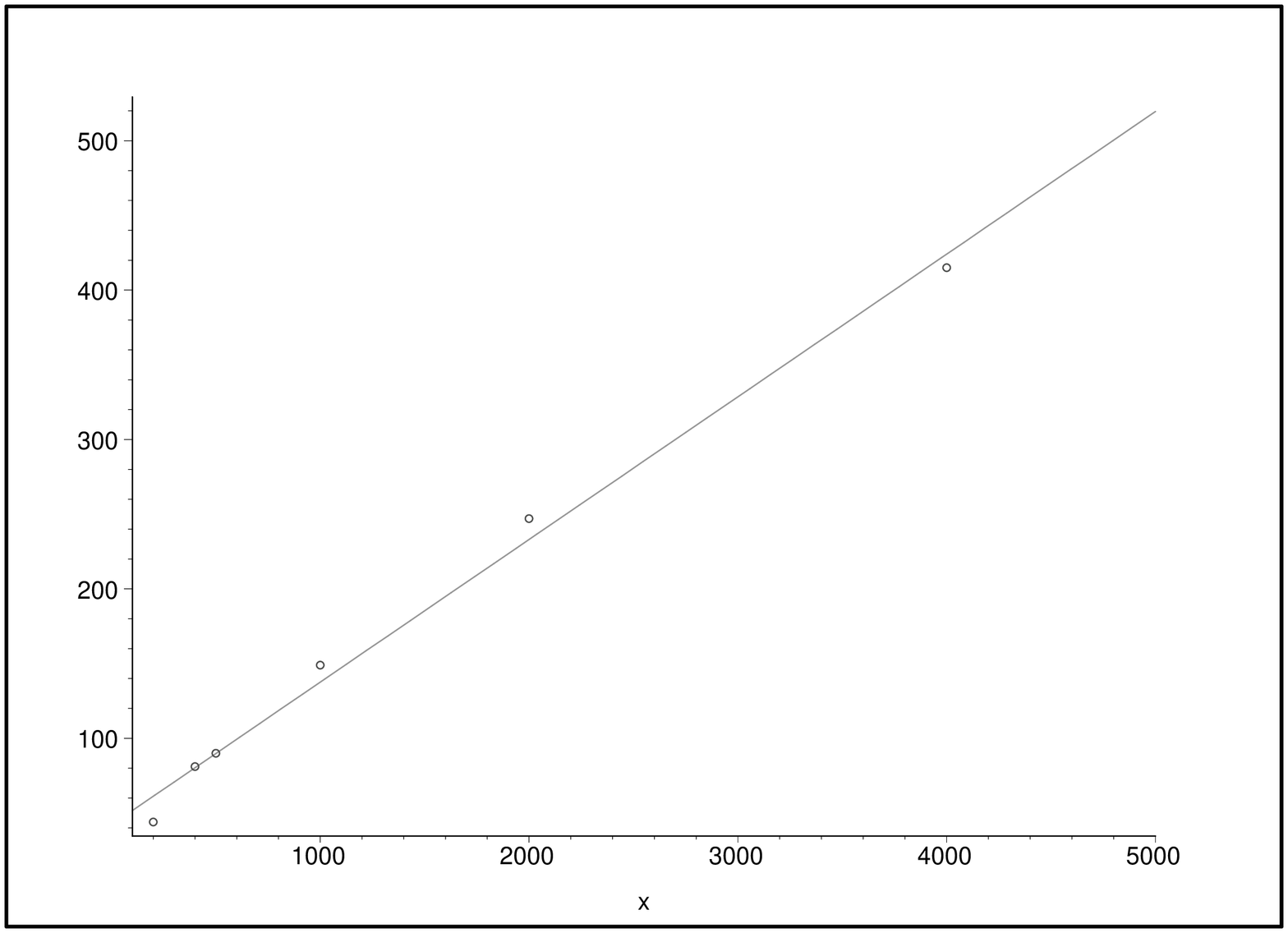}

\begin{caption}
{ Dispersions for $\nu_0 = 20$ as a function of
$x \equiv \nu_{fin}$ }
\end{caption}

\end{figure}

One can see that it can be approximately the
straight line
$$
\delta \nu = 42+ 0.095 \nu_{fin}
$$

\item
For $\nu_0 = 50$ one can see the analogous
dependence of $\delta \nu$ on $\nu_{fin}$ drawn
in Figure 2

\begin{figure}[hgh]

\includegraphics[angle=270,totalheight=8cm]{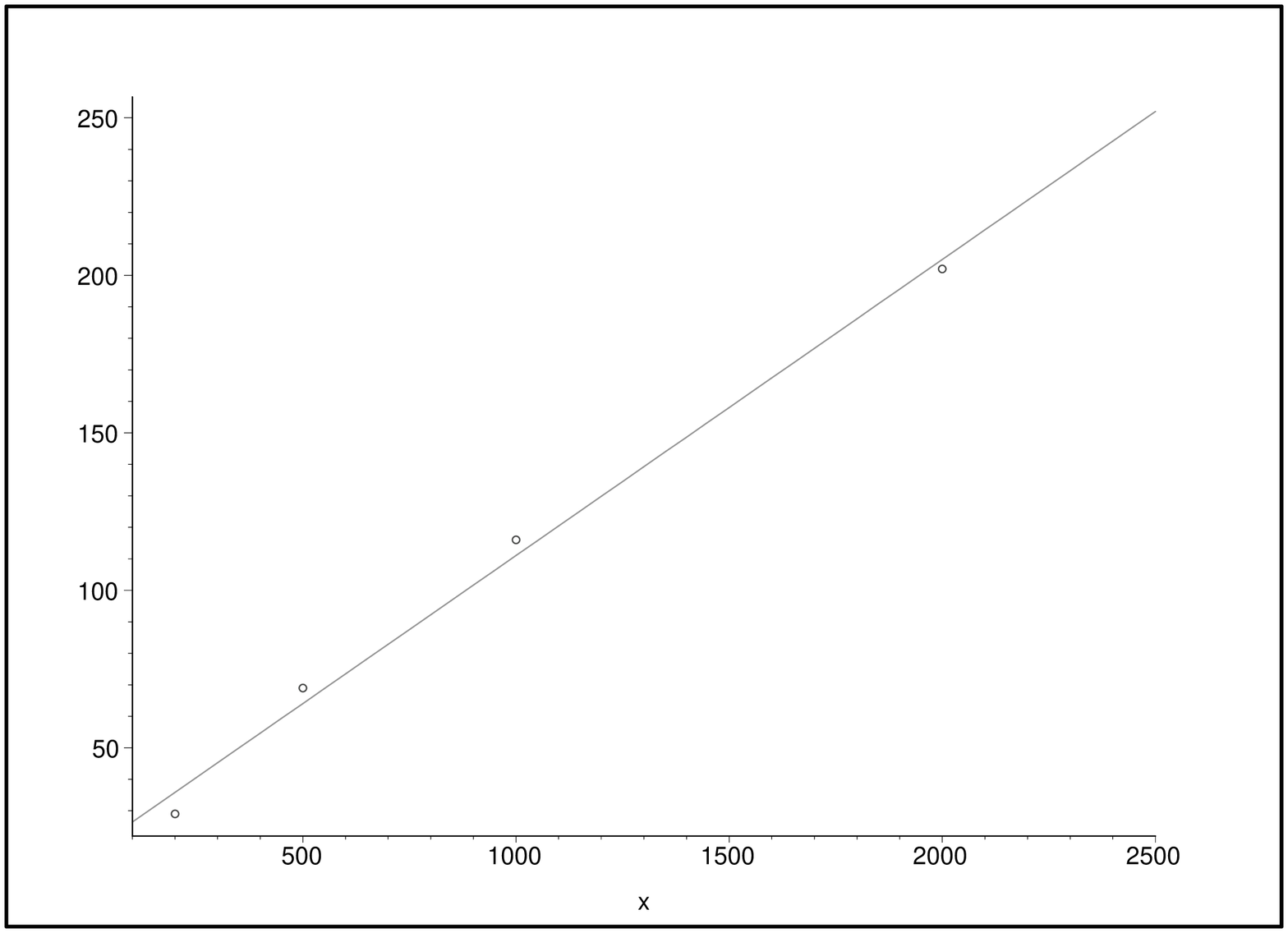}

\begin{caption}
{ Dispersions for $\nu_0 = 50$ as a function of
$x \equiv \nu_{fin}$ }
\end{caption}

\end{figure}

One can see that it is also approximately the
straight line
$$
\delta \nu = 17 + 0.094 \nu_{fin}
$$

\end{itemize}

The result of these two numerical pictures is
that one can not say that the fluctuations
disappear when $\nu$ goes to infinity. The result
is that
$$
lim_{\nu \rightarrow \infty}  \delta \nu = 0.095
\nu
$$

The direct sequence is that one has to take
fluctuation effects of growth into account.

Fortunately, the true result is a little bit more optimistic. In
reality $\delta \nu / \nu$ becomes small for enough big $\nu$. More
precisely the result of mentioned simulations give
$$
\delta \nu \sim \nu^{1-\epsilon}
$$
where
$$
\epsilon \simeq 0.22 \pm 0.05
$$

It can be seen from corresponding pictures drawn
in logarithmic scales. The following Figure 3
shows the dependence of $\ln \delta \nu$ on $\nu$
for $\nu_0 = 20$

\begin{figure}[hgh]

\includegraphics[angle=270,totalheight=8cm]{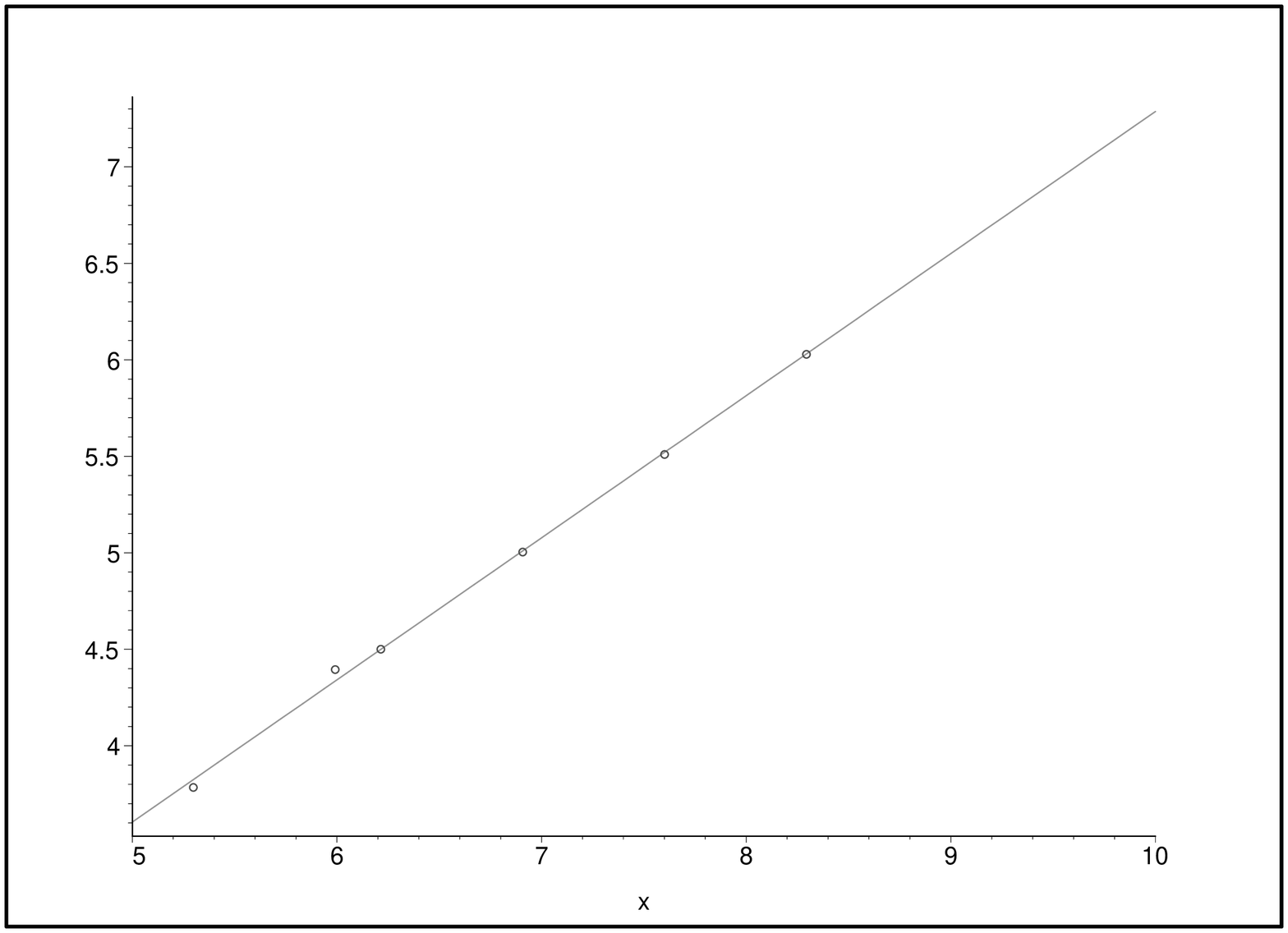}

\begin{caption}
{ Logarithm of dispersions for $\nu_0 = 20$ as a
function of $x \equiv \ln \nu_{fin}$ }
\end{caption}

\end{figure}

The linear approximation
$$
\ln \delta \nu =  0.74 \ln \nu  - 0.76
$$
is also drawn.

The same dependence for $\nu_0 = 50 $ is drawn in
Figure 4

\begin{figure}[hgh]

\includegraphics[angle=270,totalheight=8cm]{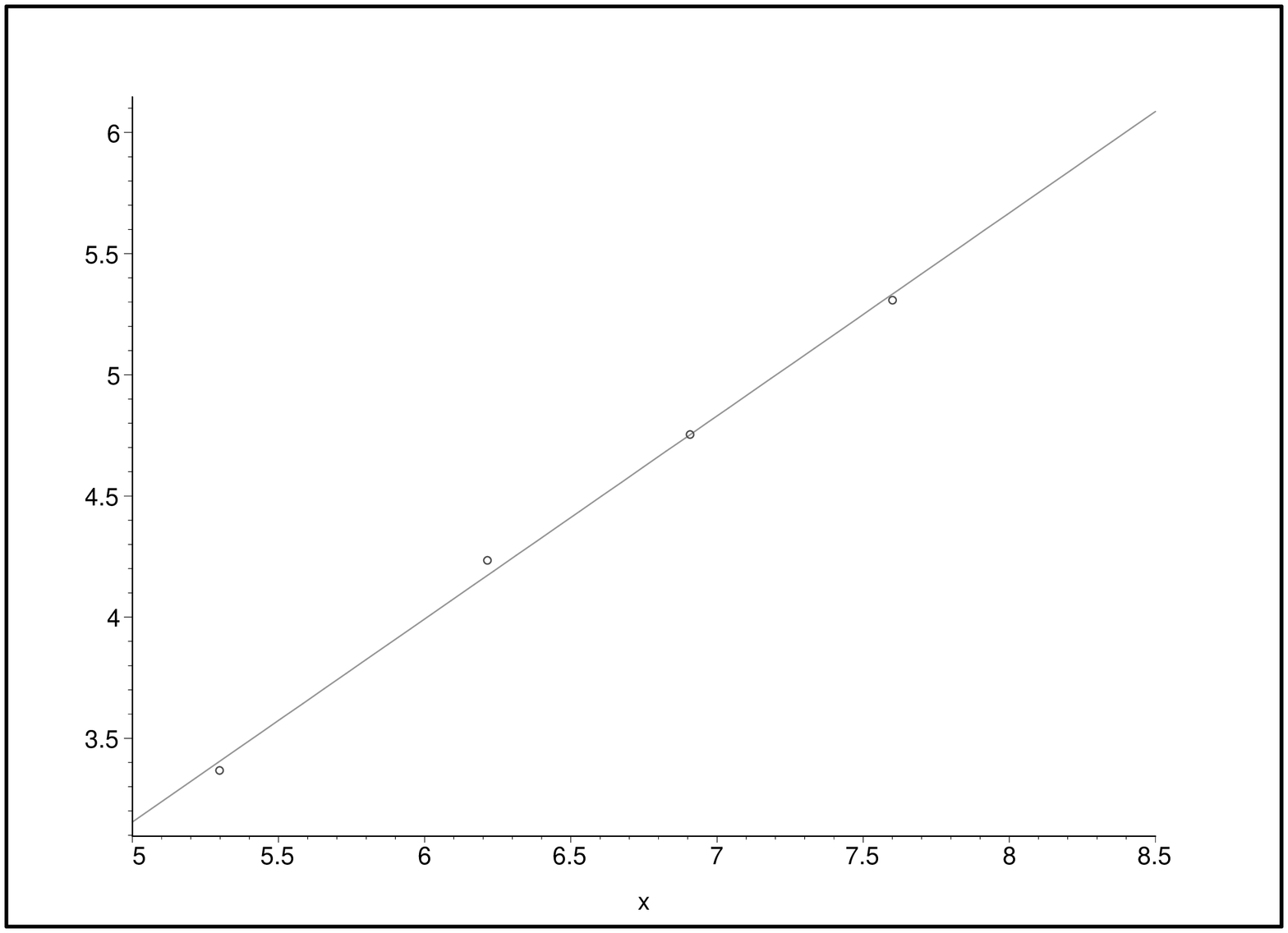}

\begin{caption}
{ Logarithm of dispersions for $\nu_0 = 50$ as a
function of $x \equiv \ln \nu_{fin}$ }
\end{caption}

\end{figure}

The best linear approximation
$$
\ln \delta \nu =  0.83 \ln \nu  - 1.03
$$
is also drawn.

So, the convergence $\delta \nu / \nu \rightarrow 0$ formally takes
place. But for  every concrete (finite) situation one has to take
this correction into account.

\section{Theoretical explanation for dependence
of dispersion on the final size}

Here we are interested in the dependence of $\delta \nu$ on
$\nu_{fin}$. This dependence will be like a power  one and the value
of the power has to be determined.

The starting point of explanation will be equation (\ref{r}). We can
see that $D_r =\frac{1}{3r^2}$ has a singularity at $r=0$.
Certainly, $r=0$ is out of range of consideration. At least $D_r$ is
a rapidly decreasing function. It means that the main blurring takes
place in initial period of time of growth (i.e. some
$\delta_{initial} r$ has been formed) and later this blurring is
mainly translated along $\nu$-scale.

It is rather easy to estimate the role of
translation. Really, due to translation
$$
\delta \nu = \delta_{initial} r \  (\frac{d\nu}{dr}) =
\delta_{initial}
 r \  3 \nu^{2/3}
$$
since the value of initial dispersion $\delta r$
is fixed we come to the power like dependence
$$
\delta \nu \sim \nu_{fin}^{2/3}
$$
The power is $2/3$.

One has also to consider the standard diffusion
of purely independent events with blurring  in
the  $r$ scale. This leads to $\delta r \sim
r^{1/2}$. Then
$$\delta \nu \sim  r^{1/2} = \nu^{1/6}$$

The power like dependence together with scaling
invariancy means that some fixed part of
evolution forms the dispersion $\delta_{initial}
r $. For this value one can take
$$\delta_{initial} \nu \sim  r_{initial}^{1/2}
= \nu^{1/6}_{initial}$$

Since $\nu_{initial} = \alpha \nu_{fin}$, where
$\alpha$ is some fixed parameter, one can come to
$$\delta_{initial} \nu \sim
 \nu^{1/6}_{fin}$$
and
$$
\delta \nu \sim \nu^{1/6}_{fin}
(\frac{d\nu}{dr}) \sim
 \nu_{fin}^{2/3+1/6} = \nu_{fin}^{0.83}
$$

Numerical simulations confirm this result.


\section{Numerical results for
dispersion as a function of the initial size}

The problem of establishing the dependence of
$\delta \nu$ is not yet solved because there
exists the uncomfortable dependence of $\delta
\nu$ on the starting size $\nu_0$.

One can not put $\nu_0$ to zero and solve this
problem because in numerical simulation one has
to exclude the dissolution of embryos.

The simulations with finite $\nu_0$ shows the
irregular behavior of $\delta \nu$ for small
$\nu$. Seeking for the power like dependence we
shall draw curves in logarithmic scales. The
dependence of $\ln \delta \nu$ on $\ln \nu_0$ is
drawn in Figure 5 for $\nu_{fin} = 1000$ and
Figure 6 for $\nu_{fin} = 2000$

\begin{figure}[hgh]

\includegraphics[angle=270,totalheight=8cm]{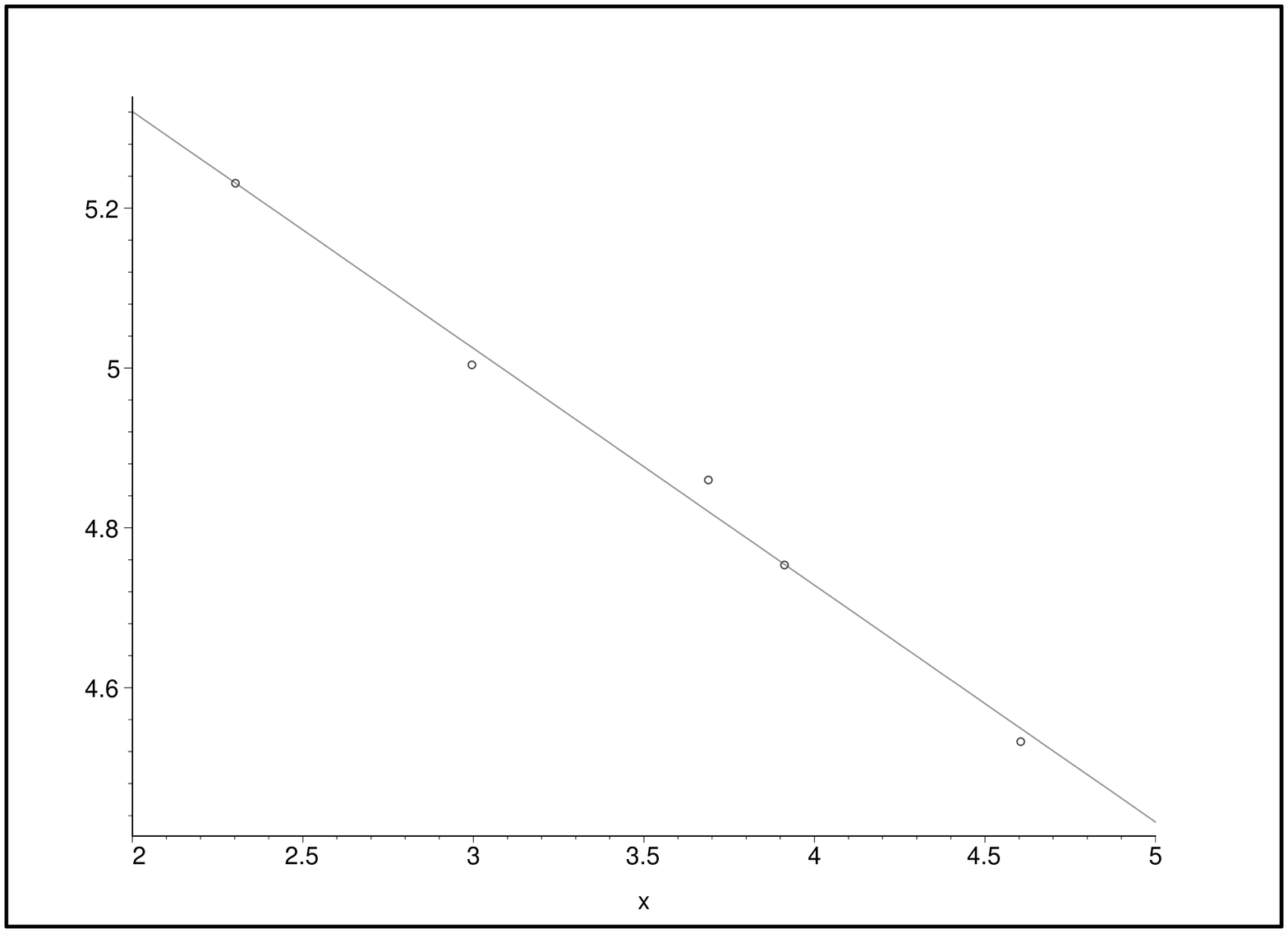}

\begin{caption}
{ Logarithm of dispersions for $\nu_{fin} = 1000$
as a function of $x \equiv \ln \nu_{0}$ }
\end{caption}

\end{figure}

\begin{figure}[hgh]

\includegraphics[angle=270,totalheight=8cm]{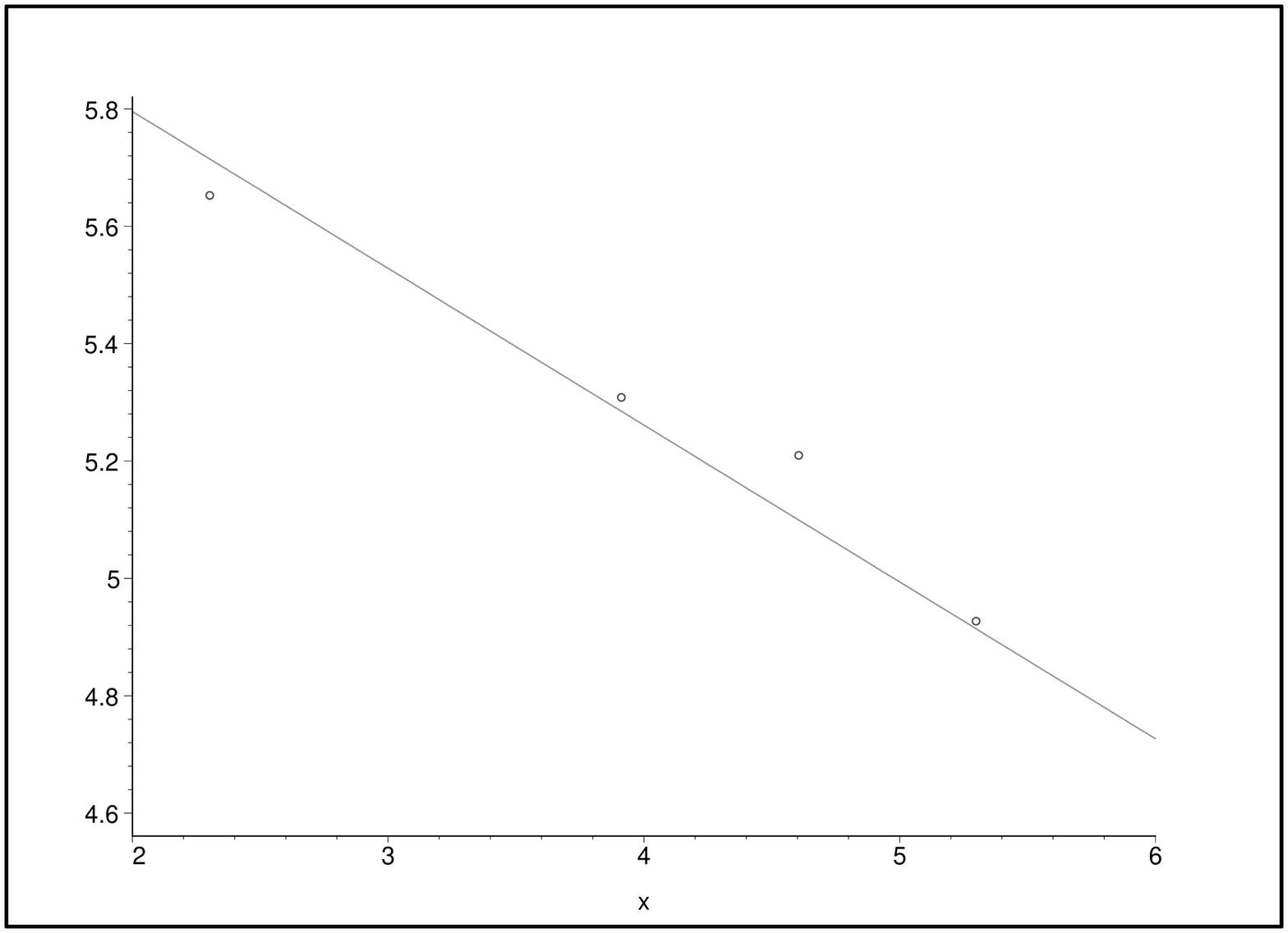}

\begin{caption}
{ Logarithm of dispersions for $\nu_{fin} = 2000$
as a function of $x \equiv \ln \nu_{0}$ }
\end{caption}

\end{figure}

The straight lines drawn in these figures are the
following
\begin{itemize}
\item
In Figure 5
$$
y = 6 -0.3 x
$$
\item
In Figure 6
$$
y = 6.3 - 0.27 x
$$
\end{itemize}

\section{Theoretical explanation of
dependence on initial size}

Again we are seeking for the power in the power
like dependence of $\delta \nu$ on $\nu_0$.
Having noticed that $D_r \sim \nu^{-2/3}$ we can
rewrite this equation as
$$
D_r \sim (t-t')^{-2}
$$
where $t'$ is the moment corresponding to $r$.

We see that $D_r$ is mainly concentrated in
initial moments of time. This allows us to state
that approximately
$$
\delta r \sim (\int D_r  dt )^{1/2}
$$

Having calculated the last integral and
interested in behavior at the lower limit of
integration we come to
$$
t^{-1/2} \sim \delta r
$$

In terms of $\nu \sim t^3$ one can rewrite the
last equation as
$$
\nu^{-1/6} \sim \delta r
$$

The numerical results confirm in general features
the small negative value of the power. But the
some discrepancy between theoretical model and
results of simulation still remains.

Then we have to  use the ideas of the
self-similarity. There is no difference between
$\nu_0$ and $\nu_{fin}$. Really, the value
$\nu_0$ is the final value  for the region
$[0,\nu_0]$ which is out of consideration. The
dispersion was supposed to be like $\nu_0^{1/2}$.
But since $\nu_0$ can be interpreted as a final
size, we know that dispersion really will be
$\nu_0^{\frac{1}{6} +\frac{2}{3}}$.

Then we have to increase  the power of the previous result in
$$
(1/6 + 1/3) / (1/2)
$$
times. This leads to
$$
\delta r \sim  \nu^{ - [(2/3+1/6)/(1/2)]*(1/6)} = \nu^{-5/18}
$$

Then the coincidence between theoretical and
experimental results becomes satisfactory.

\section{Calculations for the kernel}

Since the functional dependence on $\nu_0$ and on
$\nu_{fin}$ have been determined we don't know
only the numerical coefficient in the formula
$$
\delta \nu = const \nu_0^{-5/18} \nu_{fin}^{5/6}
$$ This coefficient can be determined
by the unique computer simulation or by
application of analytical model to one of the
concrete situations.

Here we shall show no more than an illustrative example of
calculations. The weak feature is that it does not correspond to all
approaches used above.

Let us choose the situation used in simulations
for the mean value. The value of dispersion has
to be calculated as following:
\begin{itemize}
\item
The moments of time
\begin{equation}\label{t0}
t_0=  3
\nu_0^{1/3}/\zeta
\end{equation}
 and
 \begin{equation}\label{tf}
 t_f = 3
\nu_{fin}^{1/3}/\zeta
\end{equation}
 have to be
determined.
\item
Dispersion in $r$ scale  is
$$
\delta r = \sqrt{2 N_{total}}
$$
where $N_{total}$ is the effective total number
of events. It cam be approximately calculated
according to the following formula
$$
N_{total} =  2 \int_{t_0}^{t_f} D_r dt
$$to have an analogy  with $\delta r = \sqrt{4 D t}$ in a case $D =
const$.  Having used (\ref{difco}) and keeping in mind $\delta r
\sim \sqrt{4Dt}$ we come to
$$
\delta r = \sqrt{ 4 \int_{t_0}^{t_f} \frac{1}{9} \frac{\zeta+2}{2}
\frac{3^2}{\zeta^2  t^2} dt} = \sqrt{ 2  (\zeta+2) \frac{1}{\zeta
^2} (\frac{1}{t_0} - \frac{1}{t_f})}
$$
\item
Dispersion in $\nu$ scale will be
$$\delta {\nu} = \delta {r} *3 \nu_{fin}^{2/3}$$
After substitution of (\ref{t0}) and (\ref{tf})
we come to
$$\delta {\nu} = 3 \nu_{fin}^{2/3}
\sqrt{\frac{\zeta+2}{\zeta^2}(\frac{\zeta}{3\nu_0^{1/3}} -
\frac{\zeta}{3\nu_{fin}^{1/3}} )} \sqrt{2} $$

\item
It is quite evident that  the dispersion of self
blurring
$$
\delta_0 \nu  = \sqrt{2 (\zeta+2)/\zeta
\nu_{fin}}
$$
 has not to be added. This is because we have not divided the whole
 region into two parts but integrate $D_r$ over the whole time
 interval.
\end{itemize}

The result for dispersion without self blurring
will be
$$
\delta {\nu} = 155
$$
The addition of dispersion of self blurring leads to
$$
\delta {\nu} \sim 200
$$
 It practically
coincides with the result of simulations.

As the result the dependence of $\delta \nu$ will
be
$$
\delta \nu = 155 (\frac{\nu_{fin}}{1000})^{5/6}
(\frac{20}{\nu_0})^{5/18}
$$

\section{Application for concrete systems}

{ \bf Decay }

Now we shall apply the already known formulas for
concrete systems. It is known that two typical
types of external conditions are the decay of the
metastable  state \cite{Novoj} and the so-called
dynamic conditions for condensation
\cite{PhysRevE94}.

The  global theories of evolution based on
averaged rates of nucleation (formation of
droplets)  and  droplets growth were given in
\cite{Novoj} for decay and in \cite{PhysRevE94}
for dynamic conditions.

The peculiarities of stochastic appearance of
droplets were investigated in \cite{lanl1},
\cite{lanl2} for decay and in \cite{lanl3},
\cite{lanl4} for dynamic conditions. In these
publications the rate of growth for supercritical
embryos of liquid phase is supposed to be the
regular one.

The results in description of stochastic effects
in \cite{lanl1}, \cite{lanl2}, \cite{lanl3},
\cite{lanl4} were achieved by application of
algebraic structure of the nucleation period.
This structure will be very important below.

In the situation of decay the structure  was the following one:
\begin{itemize}
\item
Until the moment when the prescribed (by recipe
of the monodisperse approximation) part of the
total (averaged total) number of droplets has
been appeared the system simply waits. This is
the first subperiod of the nucleation period.

\item
Then the second subperiod begins. In this period
the rest droplets appear. The vapor consumption
and, thus, the behavior of supersaturation is
fully governed by the growth of droplets appeared
in the first subperiod.

\item
Dispersion of the number of droplets appeared in
the first subperiod equals to zero, the system is
simply waiting for the appearance of the
necessary amount of droplets.

\item
Dispersion of the rest droplets, i.e. of the
number of droplets appeared in the second
subperiod equals to the standard dispersion. But
due to the fact that the number of the rest
droplets is less than the total number of
droplets the final dispersion will be less than
the standard one.
\end{itemize}

The  number of droplets appeared in the first
subperiod will be marked by $N_{first}$. The
number of droplets appeared in the second
subperiod will be marked by $N_{rest}$.

All conclusions except the last one remain valid
also under the stochastic model for the droplets
growth. The last conclusion has to be
reconsidered. Really, earlier the cut off by the
vapor consumption was absolutely regular since
the number of droplets in the first subperiod has
to be absolutely fixed and, hence, the action of
these droplets was absolutely prescribed.

Now the situation is changed. The cut off of the
rest part of spectrum can be initiated in
different moments of time since the droplets from
the first part of spectrum (let us call them as
the the leading droplets) can grow faster or
slower.

One can illustrate the effect of the different
velocity of growth by the following  picture
\begin{itemize}
\item
Suppose that the droplets grow slower than the
averaged velocity. Then the moment of the cut off
will be later and more droplets than the standard
value appear in the rest part of spectrum.
\item
Suppose that the droplets grow faster than the
averaged velocity. Then the moment of the cut off
will be earlier and less droplets than the
standard  value appear in the rest part of
spectrum.
\end{itemize}

We shall call this picture as the "unbalanced
effects of growth" (UEG)

From the effective monodisperse approximation
\cite{lanl2} we know the "initial" size and the
final "size". The number of droplets in the
monodisperse peak in also known.

Certainly, the  picture  of UEG isn't absolutely
correct. It is clear that the balancing factor
appear. This factor  is similar to that which
leads to the absence of dispersion in formation
of the leading droplets.

Suppose that the initially determined leading
droplets grow slower than ordinary. Does it mean
that only these droplets have to be taken into
account in vapor consumption during the
nucleation period? Certainly not. We have simply
to include into the group of leading droplets
some new droplets appeared a few moments later
than  the initial boundary between the leading
droplets and the rest droplets. So, the existence
of a  balancing force is evident here.

Now suppose that the initially determined leading
droplets grow faster than ordinary.
 Then we have  to exclude from
the group of leading droplets some the last
droplets appeared just before
 the initial boundary between
the leading droplets and the rest droplets. So,
the balancing force here also takes place.

It is clear that the picture with unbalanced
effects of growth will estimate the stochastic
effects of growth from above. We shall use this
property to calculate the stochastic effects in
this picture and then to see that they are small
enough.

In reality the result from stochastic growth will
be some  times smaller.

Now we turn to the general case of asymptotic of
the big number of molecules in critical embryo.

To grasp the stochastic effects from above we use
the picture of unbalanced effects of growth and
calculate the effects from the stochastic
appearance of droplets
 and from
stochastic growth into dispersion. We suppose
that one can add dispersions caused by both
stochastic causes. It is true at least when one
of dispersions is much smaller than another.
Namely this situation will take place.

Dispersion of appearance is
$$
D_{appear} \sim \sqrt{2 N_{rest}}
$$
and dispersion of growth will be connected with
the fact that the number of the rest droplets
 $N_{rest}$ will fluctuate.

If we  forget that $N_{rest}$ is the argument of the square root
which  smoothes the result  we will get the estimate from above. So,
we will forget.

So, we will calculate the dispersion $ \delta_1 $
of one droplet caused by stochastic growth from
the size $\nu_{st}$ up to the final size.

Fortunately the initial size is well determined in the advanced
monodisperse approximation and it is not zero. In the previous
initial monodisperse approximation it was zero and the consideration
of stochastic effects required to give in \cite{lanl2} a new version
where the initial position of monodisperse spectrum (the $z$
-coordinate) in renormalized scale is $0.4-0.336$. Namely the new
version has to be used.

So, the value of $\nu_0$ for all droplets in the
monodisperse spectrum will be
$$
\nu_0^{1/3} = (0.4 -0.336) \nu_{fin}^{1/3}
$$

One can also determine $\nu_0$ in a following
simple manner. The actual height of monodisperse
spectrum is the unperturbed rate of nucleation
(of appearance) $J_0$. Since
$$
N_{first} \approx
J_0 \delta r
$$
 we can get the characteristic
width of the spectrum $\delta r$. This value will
be initial value for $\nu_0^{1/3}$. It is clear
that approximately this way leads to the same
value of $\nu_0$. Then since the dependence on
$\nu_0$ is like $\nu_0^{-1/6}$ we see that the
results of two ways to determine $\nu_0$ will be
approximately equivalent.

The zero value of $\nu_{0}$ would lead to
divergence, now such a danger is over and one can
see that the asymptotic at $\nu \rightarrow
\infty$ for characteristic relative deviation in
$\nu$ is
$$
\delta \nu  = \nu_{fin}^{5/6}/\nu_{fin}^{5/18}
$$
and goes to zero. The total characteristic relative deviation will
be
\begin{itemize}
\item
for one droplet in the monodisperse peak
$$
D_{unbal} \sim  \nu_{fin}^{5/6}/\nu_{fin}^{5/18}
$$
\item
for all droplets in the monodisperse peak
$$
D_{unbal} \sim  \nu_{fin}^{5/6}/\nu_{fin}^{5/18}
* 1/\sqrt{N_{first}}
$$
\end{itemize}

It will be many times less than the relative dispersion
$$
D_{stoch}  \sim 1/\sqrt{N_{rest}} \sim
1/\sqrt{N_{first}}
$$
of the number of droplets caused by stochastic
appearance.

We have to note  the absence of the shift of the averaged value of
the final size from the value calculated on the base of the regular
law of growth. The order of this  deviation is less than the
characteristic attained value. If this shift was essential the
action on the averaged number of droplets would be significant.

One has to note that the in last remark the regular law means the
regular discrete law and the main effects will be the effects of
discrete character of the droplets growth.

{\bf Dynamic conditions }

In dynamic conditions all previous arguments are
valid. The initial size $\nu_0$ has to determined
by the following way. Since in the approximation
of monodisperse consumption \cite{lanl4} the
monodisperse spectrum is formed already at
$z=-3/c$ one can say that it is formed at the
ideal conditions (the real supersaturation
coincides with the ideal one). Then the
characteristic width of spectrum will be
$c^{-1}$. Then  the starting value value
$\nu_0^{1/3}$ will have the order $c^{-1}$
$$
\nu_0^{1/3} \sim c^{-1}
$$
The final value $\nu_{fin}^{1/3} $ will have the
order $(3 \div 4) c^{-1}$
$$
\nu_{fin}^{1/3} \sim (3 \div 4) c^{-1}
$$
 All other constructions
are absolutely analogous.

Certainly, the numerical simulation can not give
correction terms where the effects of stochastic
growth can be noticed because these effects will
be corrections for corrections to the main
result. The corrections connected with the
discrete growth by steps of absorption of one
molecule will be more significant.

One has also to note that at the further periods
of evolution (after the end of nucleation) the
stochastic effects of growth will be more
important. This consideration will be presented
in a separate publication.

\end{document}